\documentclass[]{aa}
\usepackage{graphicx}
\usepackage{subfigure}
\usepackage{txfonts}

\begin{document}

\title{Response of the solar atmosphere to magnetic field evolution \\ in a coronal hole region}

\author{S. H. Yang \inst{1}
        \and
        J. Zhang \inst{1} \and C. L. Jin \inst{1} \and L. P. Li \inst{1} \and H. Y.
        Duan \inst{2}
        }

\institute{National Astronomical Observatories, Chinese Academy of
           Sciences, Beijing 100012, China\\
           \email{shuhongyang@ourstar.bao.ac.cn, zjun@ourstar.bao.ac.cn}
           \and
           Jiangsu Sopo Corporation Group Ltd., Zhenjiang 212006,
           China
           }

\titlerunning{Atmospheric response to magnetic field evolution}
\authorrunning{S. H. Yang et al.}

\date{Received 15 July 2008 / Accepted 14 April 2009}

\offprints{Shuhong Yang}


\abstract
{Coronal holes (CHs) are deemed to be the sources of the fast
solar wind streams that lead to recurrent geomagnetic storms and
have been intensively investigated, but not all the properties of
them are known well.}
{We mainly research the response of the solar atmosphere to the
magnetic field evolution in a CH region, such as magnetic flux
emergence and cancellation for both network (NT) and intranetwork
(IN).}
{We study an equatorial CH observed simultaneously by
\emph{HINODE} and \emph{STEREO} on July 27, 2007. The
\emph{HINODE}$/$SP maps are adopted to derive the physical
parameters of the photosphere and to research the magnetic field
evolution and distribution. The G band and \ion{Ca}{ii} H images
with high tempo-spatial resolution from \emph{HINODE}$/$BFI and
the multi-wavelength data from \emph{STEREO}$/$EUVI are utilized
to study the corresponding atmospheric response of different
overlying layers.}
{We explore an emerging dipole locating at the CH boundary.
Mini-scale arch filaments (AFs) accompanying the emerging dipole
were observed with the \ion{Ca}{ii} H line. During the separation
of the dipolar footpoints, three AFs appeared and expanded in
turn. The first AF divided into two segments in its late stage,
while the second and third AFs erupted in their late stages. The
lifetimes of these three AFs are 4, 6, 10 minutes, and the two
intervals between the three divisions or eruptions are 18 and 12
minutes, respectively. We display an example of mixed-polarity
flux emergence of IN fields within the CH and present the
corresponding chromospheric response. With the increase of the
integrated magnetic flux, the brightness of the \ion{Ca}{ii} H
images exhibits an increasing trend. We also study magnetic flux
cancellations of NT fields locating at the CH boundary and present
the obvious chromospheric and coronal response. We notice that the
brighter regions seen in the 171 {\AA} images are relevant to the
interacting magnetic elements. By examining the magnetic NT and IN
elements and the response of different atmospheric layers, we
obtain good positive linear correlations between the NT magnetic
flux densities and the brightness of both G band (correlation
coefficient 0.85) and \ion{Ca}{ii} H (correlation coefficient
0.58).}
{}


\maketitle
%

\section{Introduction}

Coronal holes (CHs) are regions on the Sun where the magnetic
fields are dominated by one magnetic polarity and the magnetic
lines are open to interplanetary space (Bohlin 1977). These
unipolar structures are presumed to originate at about 0.7
R$_{\odot}$ distance from the solar center (Stepanian 1995). CHs
were identified as early as in 1951 by Waldmeier (1951) from the
earliest photo of the solar corona. Observed with X-ray (Underwood
\& Muney 1967) and EUV line (Reeves \& Parkinson 1970), they
appear as dark and void areas due to their lower densities and
lower temperatures compared with that of the quiet Sun (QS) (Munro
\& Withbroe 1972; Harvey 1996). But if observed with \ion{He}{i}
10830 {\AA} line, they are brighter areas (Zirker 1977; Harvey \&
Sheeley 1979). CHs are usually classified into three different
categories according to their locations and lifetimes: polar,
non-polar (isolated), and transient (Harvey \& Recely 2002). Polar
CHs can persist for about several years, non-polar ones always
many solar rotations, while transient ones only several days. They
are deemed to be the sources of the fast solar wind streams that
lead to recurrent geomagnetic storms (Krieger et al. 1973;
Tsurutani \& Gonzalez 1987; Crooker \& Cliver 1994; Cranmer 2002;
Tu et al. 2005).

A mass of research concerning CHs has been extensively done in
recent years. These studies refer nearly all the properties of
CHs, such as positional distribution and periodic variation
(Belenko 2001; Maravilla et al. 2001; Harvey \& Recely 2002;
Bilenko 2002, 2004; Hofer \& Storini 2002; Mahajan et al. 2002;
Temmer et al. 2007), magnetic field structure (Meunier 2005;
Wiegelmann et al. 2005; Tian et al. 2008), magnetic field
evolution (Wang \& Sheeley 2004; Yamauchi et al. 2004; Zhang et
al. 2006), temperature variation (Patsourakos et al. 2002; Wilhelm
2006; Zhang et al. 2007), element abundance (Feldman \& Laming
2000; Laming \& Feldman 2003), Doppler velocity (Raju et al. 2000;
Stucki et al. 2000, 2002; Jordan et al. 2001; Wilhelm et al. 2002;
Kobanov et al. 2003; Xia et al. 2003, 2004; Akinari 2007;
Teplitskaya et al. 2007), fast solar wind (Giordano et al. 2000;
Hackenberg et al. 2000; Patsourakos \& Vial 2000; Wilhelm et al.
2000; Teriaca et al. 2003; Zhang et al. 2002, 2003, 2005a; Janse
et al. 2007) and wave in CH (Moran 2003; Zhang 2003; Markovskii \&
Hollweg 2004; O'Shea et al. 2005, 2006, 2007; Zhang et al. 2005b;
Dwivedi et al. 2006; Kobanov \& Sklya 2007; Srivastava et al.
2007; Wu et al. 2007).

According to previous studies, CHs are not absolutely unipolar and
there exist many closed coronal loops (eg. Levine 1977; Zhang et
al. 2006). The magnetic network (NT; Leighton et al. 1962)
elements are believed to be consisted of both low-lying loops and
large-scale open magnetic funnels (Dowdy et al. 1986; Dowdy 1993),
while the intranetwork (IN; Livingston \& Harvey 1975) ones only
contribute low-lying loops (Wiegelmann \& Solanki 2004; Wiegelmann
et al. 2005). In equatorial CHs, the chromosphere and transition
region are highly structured. Their structures are similar to
those in the QS region (Warren \& Winebarger 2000; Feldman et al.
2001). Above the upper transition region, if observed with X-ray
or EUV line, most of the structures disappear and the corona
becomes much darker and more homogeneous than in the QS, except
for some bright points (Xia et al. 2004).

Magnetic flux emergence and cancellation are the main forms of
magnetic field evolution in the Sun (Zhang et al. 1998a$-$c).
In an emerging flux region, new emerging dipoles may lead to the
formation of arch filament systems (AFSs). AFSs were first studied
by Bruzek (1967), whereafter their properties, such as size,
shape, lifetime, evolution, structure and so on, have been
extensively investigated (Bruzek 1969; Weart 1970; Frazier 1972;
Chou \& Zirin 1988; Alissandrakis et al. 1990; Georgakilas et al.
1990; Tsiropoula et al. 1992; Chou 1993).
The
reconnection occurring at CH boundaries is crucial to the
evolution and rigid rotation of CHs (Wang et al. 1996; Kahler \&
Hudson 2002).

There is always a good correlation between the QS magnetic NT
elements and the chromospheric structures since the magnetic NT
features are always bright in the \ion{Ca}{ii} line (Leighton
1959; Frazier 1970; Skumanich et al. 1975; Stenflo \& Harvey 1985;
Zirin 1988; Rezaei et al. 2007). Although the studies from
Sivaraman \& Livingston (1982) and Sivaraman et al. (2000) reveal
that magnetic concentrations play a major role in the formation of
\ion{Ca}{ii} K line IN elements and there is a one-to-one
relationship between K line IN bright elements and magnetic
features, some other authors (eg. Remling et al. 1996; Steffens et
al. 1996) report that there is no correlation between small-scale
magnetic elements and locations of bright Ca K$_{2\upsilon}$ IN
elements. Nindos \& Zirin (1998) studied quantitatively the
relation between the intensity of \ion{Ca}{ii} K line bright
features and the intensity of the associated magnetic elements in
the QS. They found that there is an almost linear correlation
between the K-line intensities and the absolute values of the
magnetic field strength for the stronger NT elements, while this
correlation disappears for the weaker magnetic elements. Lites et
al. (1999) investigated the QS IN and found no direct correlation
between the presence of magnetic features with apparent flux
density above 3 Mx cm$^{-2}$ and the occurrence of H$_{2\upsilon}$
brightenings. They also found no correspondence between
H$_{2\upsilon}$ grains and the horizontal-field IN features. Then
Worden et al. (1999) found only a random correspondence between
bright cell grains and regions of IN magnetic flux as seen in
\ion{H}{i} Ly$\alpha$ and 160 nm lines. Recently, Rezaei et al.
(2007) investigated the relationship between the photospheric
magnetic field and the emission of the mid-chromosphere and found
that the emission in the NT is correlated with the magnetic flux
density while there is no correlation between the integrated
emission of Ca core-line and the magnetic flux density in the IN.

Owing to the restriction of observations, not all the
characteristics of CHs are known well. Several newly launched
space-based instruments, especially \emph{HINODE} (Kosugi et al.
2007), have provided unprecedentedly high spatial and temporal
resolution data, which is just the extremely important factor of
investigating the fine structures in detail. Furthermore,
multi-passband images from the Solar Terrestrial Relations
Observatory (\emph{STEREO}; Howard et al. 2008; Kaiser et al.
2008) give us plentiful information about the solar atmosphere.
These data can be used together to study the magnetic field
evolution and the response of overlying atmosphere, which will be
helpful for us to understand the physical mechanism of the coronal
heating and the solar wind acceleration.

In order to study magnetic flux emergence, cancellation,
distribution, and the atmospheric response of different overlying
layers in CHs, we investigate an equatorial CH of predominantly
negative polarity in this paper. In Sect. 2, we introduce the
observations and data analysis. In Sect. 3, we give some typical
examples of small-scale magnetic field evolution and the
corresponding atmospheric response, as well as the relationship
between the magnetic intensities and the brightness of overlying
solar atmosphere. The conclusions and discussion are presented in
Sect. 4.


\section{Observations and data analysis}

\begin{table}
\caption{Data sets mainly used in this study.} 
\label{table1} 
\centering 
\begin{tabular}{c c c c c} 
\hline\hline 
Data Set & Observation & Cadence & Pixel Size & FOV \\ 
 & & (minutes) & (arcsec)& (arcsec$^{2}$) \\
\hline 
I  & HINODE$/$SP & 32$^{\mathrm{a}}$ & 0.32 & 151.14${\times}$162.30 \\ 
II & HINODE$/$Gband & 2 & 0.11 & 223.15${\times}$111.58 \\
   & HINODE$/$\ion{Ca}{ii} H    & 2 & 0.11 & 223.15${\times}$111.58 \\
III & STEREO$/$304 & 10 & 1.59 & full disk \\
    & STEREO$/$171 & 2.5 & 1.59 & full disk \\
    & STEREO$/$195 & 10 & 1.59 & full disk \\
    & STEREO$/$284 & 20 & 1.59 & full disk\\
\hline 
\end{tabular}
\begin{list}{}{}
\item[$^{\mathrm{a}}$] Scan time for one SP map.
\end{list}
\end{table}

\begin{figure}
\centering
\resizebox{8.9cm}{!} {\includegraphics [scale=.5]{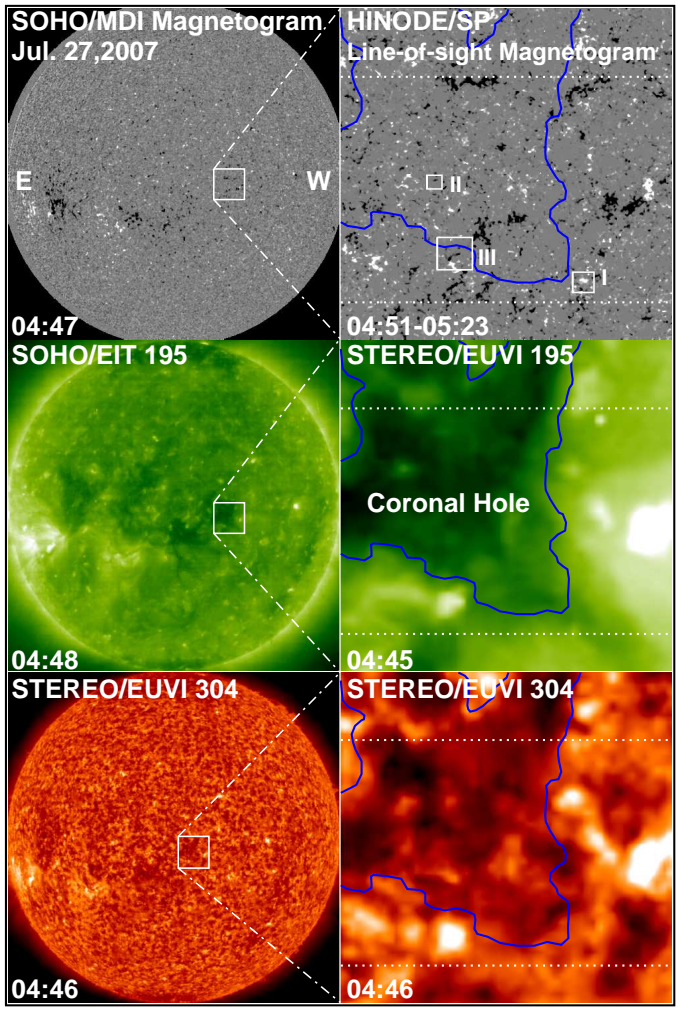}}
\caption{\emph{SOHO}$/$MDI full-disk magnetogram (top left),
\emph{SOHO}$/$EIT 195 {\AA} image (middle left), and
\emph{STEREO}$/$EUVI 304 {\AA} image (bottom left). Three windows
in the three full-disk images outline the FOV of
\emph{HINODE}$/$SP magnetograms. The right column shows a
line-of-sight magnetogram from \emph{HINODE}$/$SP, a 195 {\AA}
image and a 304 {\AA} image from \emph{STEREO} from top to bottom.
The magnetograms from MDI and SP are displayed with (-50 G, 50 G)
and (-80 G, 80 G) scales, respectively. The blue curves delineate
the CH boundary derived from the \emph{STEREO}$/$EUVI 284 {\AA}
image obtained at 04:46 UT, while the dotted lines indicate the
FOV of the G band and \ion{Ca}{ii} H images. The rectangles in the
top-right image delineate the regions (``I", ``II", ``III") that
are specially studied.\label{fig1}}
\end{figure}

\begin{figure}
\centering
\resizebox{8.9cm}{!}{\includegraphics[bb=95 237 457
563,clip,angle=0]{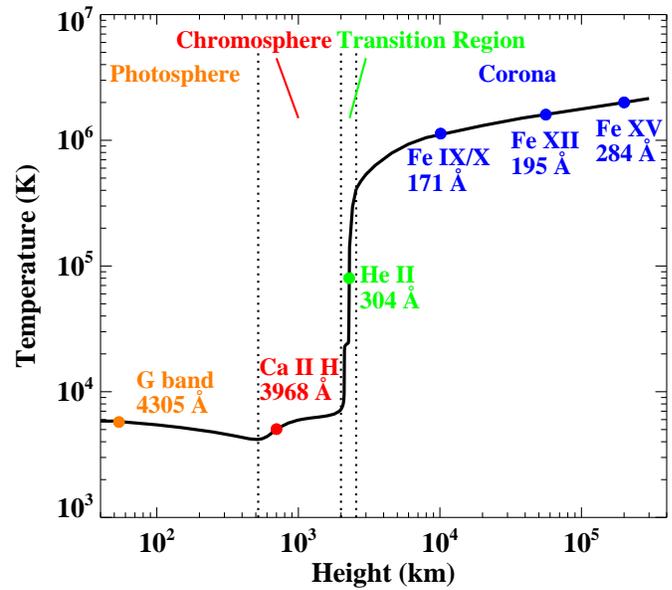}} \caption{Variation of the temperature
versus the height above the $\tau_{5000}$=1 (optical depth unity
in the continuum at 5000 {\AA}) surface. The G band and
\ion{Ca}{ii} H lines are emitted respectively in the lower
photosphere and lower chromosphere, and their mean formation
heights (54 km and 700 km, respectively) are all marked with
filled circles in this figure. The peak formation temperatures of
\ion{He}{ii} (304 {\AA}) line emitted in the transition region,
and \ion{Fe}{ix$/$x} (171 {\AA}), \ion{Fe}{xii} (195 {\AA}),
\ion{Fe}{xv} (284 {\AA}) lines which are all emitted in the inner
solar corona are 80{\,}000 K, 1.3 MK, 1.6 MK, 2.0 MK,
respectively. They are also marked with filled circles in this
figure.\label{fig2}}
\end{figure}

The observations were carried out on July 27, 2007, using the
Solar Optical Telescope (SOT; Ichimoto et al. 2008; Shimizu et al.
2008; Suematsu et al. 2008; Tsuneta et al. 2008) instrument on
board \emph{HINODE}, the Extreme Ultra Violet Imager (EUVI; Howard
et al. 2008) telescope of the Sun-Earth Connection Coronal and
Heliospheric Investigation (SECCHI; Howard et al. 2008) instrument
aboard \emph{STEREO}, the Michelson Doppler Imager (MDI; Scherrer
et al.1995) and Extreme-ultraviolet Imaging Telescope (EIT;
Delaboudini\`{e}re et al. 1995) onboard the Solar and Heliospheric
Observatory (\emph{SOHO}; Domingo et al. 1995).

The data sets mainly used in this study are summarized in Table 1.
The Spectro-Polarimeter (SP; Lites et al. 2001) in
\emph{HINODE}$/$SOT provides observations in four modes: normal,
fast, dynamics and deep maps. Five sets of \emph{HINODE}$/$SP maps
adopted here were observed in fast map mode between 01:36 UT and
05:58 UT.
They are all centered at about
11${^\circ}$W and 1${^\circ}$S and cover a large fraction of a CH
area. Each map is
constructed from 512${\times}$512 Stokes (I, Q, U and V) profiles
of the photospheric \ion{Fe}{i} 6301.5 {\AA} and 6302.5 {\AA}
lines with a spectral sampling of 21.5 m{\AA}. The noise level in
the continuum polarization is about 1.5$\times$10$^{-3}$ I$_{c}$.
For the observation of each spectrograph slit, the integrated
exposure time is 3.2 s. The scan direction is along the east-west
direction with a scan step of 0{\arcsec}.295.

By using the Stokes spectrum inversion code based on the
assumption of Milne-Eddington atmospheres (Yokoyama, T. private
communication), we have successfully derived from the raw data
numerous physical parameters, such as the three components of
magnetic fields (the field strength \emph{B}, the inclination
angle $\gamma$, and the azimuth angle $\phi$), the stray light
fraction $\alpha$, and the Doppler velocity \emph{V}$_{los}$.
Here, $\gamma$ is the angle between the vector magnetic field
\emph{\textbf{B}} and the LOS direction, and $\phi$ is the angle
from the east$-$west direction to the projection of
\emph{\textbf{B}} on the plane perpendicular to the LOS direction.
Then the vector magnetic field \emph{\textbf{B}} is shown by the
LOS field (1$-$$\alpha$)\emph{B}$\cos$$\gamma$ and the transverse
field (1$-$$\alpha$)$^{1/2}$\emph{B}$\sin$$\gamma$. The Doppler
shifts are derived from the \ion{Fe}{i} 6302.5 {\AA} Stokes I
profiles and averaged over the whole field of view (FOV) of the SP
maps.

Also employed from \emph{HINODE} are G band and \ion{Ca}{ii} H
images obtained by the Broadband Filter Imager (BFI; Kosugi et al.
2007).
Their FOV only covers part of the SP maps (shown with the dotted
lines in Fig. 1). Moreover, we adopt the EUVI data from satellite
A of \emph{STEREO}. The EUVI telescope observed the full-Sun in
four spectral channels (304 {\AA}, 171 {\AA}, 195 {\AA} and 284
{\AA}).
The
\emph{HINODE}$/$BFI and \emph{STEREO}$/$EUVI images used here are
selected according to the scanning time of the SP maps.

The images are all prepared by applying the standard processing
routines, including flat field correction, dark current and
pedestal subtraction, bad camera pixel correction, camera readout
error correction, cosmic ray removal, et al.. Then we co-align all
the images and SP maps carefully. Since both G band and
\ion{Ca}{ii} H observations were taken with fixed pointing, first
we co-align these images to each other. The bright features of
\ion{Ca}{ii} H line images are highly coincident with the
underlying magnetic field features both in location and shape
especially for NT elements (Warren \& Winebarger 2000; Feldman et
al. 2001; Xia et al. 2004), so we use the bright points to
co-align the G band and \ion{Ca}{ii} H images with the LOS
magnetograms. For the co-alignment between the \emph{STEREO}
images and the SP maps, we look to the \emph{SOHO}$/$MDI and
\emph{SOHO}$/$EIT images for help. At first, the LOS magnetograms
(SP maps) from \emph{HINODE} are co-aligned to the MDI full-disk
magnetograms. Since the EIT and MDI images are from the same
satellite and they can be co-aligned easily, so the SP maps can be
co-aligned to the EIT images with no difficulty. Then the
\emph{STEREO} 195 {\AA} images are co-aligned to the \emph{SOHO}
195 {\AA} images according to their bright features. After that,
the \emph{STEREO} 195 {\AA} images (also 304 {\AA}, 171 {\AA} and
284 {\AA}) have been co-aligned with the SP maps.

We determine the CH boundary with the brightness gradient method
developed by Shen et al. (2006). In an EUV 284 {\AA} image, each
pixel has its own recorded brightness, \emph{b}. For any given
value of \emph{b}, we can plot the contour and then calculate the
area, \emph{A}, enclosed by each contour. The derivation
\emph{f}=$\delta$\emph{b}$/$$\delta$\emph{A} is used to determine
the boundary of a CH. The CH boundary is at the place where
\emph{f}=\emph{f}$_{max}$.

Figure 1 shows a \emph{SOHO}$/$MDI magnetogram (top left), a
\emph{SOHO}$/$EIT 195 {\AA} image (middle left) and a
\emph{STEREO}$/$EUVI 304 {\AA} image (bottom left). Three windows
in the three full-disk images outline the FOV of
\emph{HINODE}$/$SP magnetograms. The right column shows
respectively a LOS magnetogram from \emph{HINODE}, a 195 {\AA}
image and a 304 {\AA} image from \emph{STEREO}. In order to
improve the image contrast to visualize the fine structures
better, a log-logarithmic scale is applied to the \emph{STEREO}
images presented in all the figures.

Figure 2 shows the typical variation of the temperature versus the
height of the solar atmosphere above the $\tau_{5000}$=1 surface
(Reeves et al. 1977; Vernazza et al. 1981). The G band and
\ion{Ca}{ii} H lines are emitted respectively in the photosphere
and lower chromosphere, and their mean formation heights are 54 km
(Carlsson et al. 2004) and 700 km (Beck et al. 2008),
respectively. The \ion{He}{ii} (304 {\AA}) line is emitted in the
transition region, while \ion{Fe}{ix$/$x} (171 {\AA}),
\ion{Fe}{xii} (195 {\AA}), \ion{Fe}{xv} (284 {\AA}) lines are all
formed in the inner solar corona. The peak formation temperatures
of these four passbands are 80{\,}000 K, 1.3 MK, 1.6 MK, 2.0 MK,
respectively (Delaboudini\`{e}re et al. 1995). All these mean
heights or peak temperatures are marked with filled circles in
this figure. We can see clearly that these different lines (or
images) represent different heights (or layers) of the solar
atmosphere. This means that, in order to study the response of the
different overlying layers to the magnetic field evolution and
distribution, we can study the structures seen in different
pass-band images.

It should be mentioned that all the spectral images are separately
composed into integrated new images with time slices synchronised
with the time sequences of SP scanning, so as to basically
maintain their temporal correspondence.


\section{Results}
\subsection{Magnetic flux emergence at the CH boundary}

\begin{figure}
\centering
\resizebox{8.9cm}{!}{\subfigure{\includegraphics
[scale=.5]{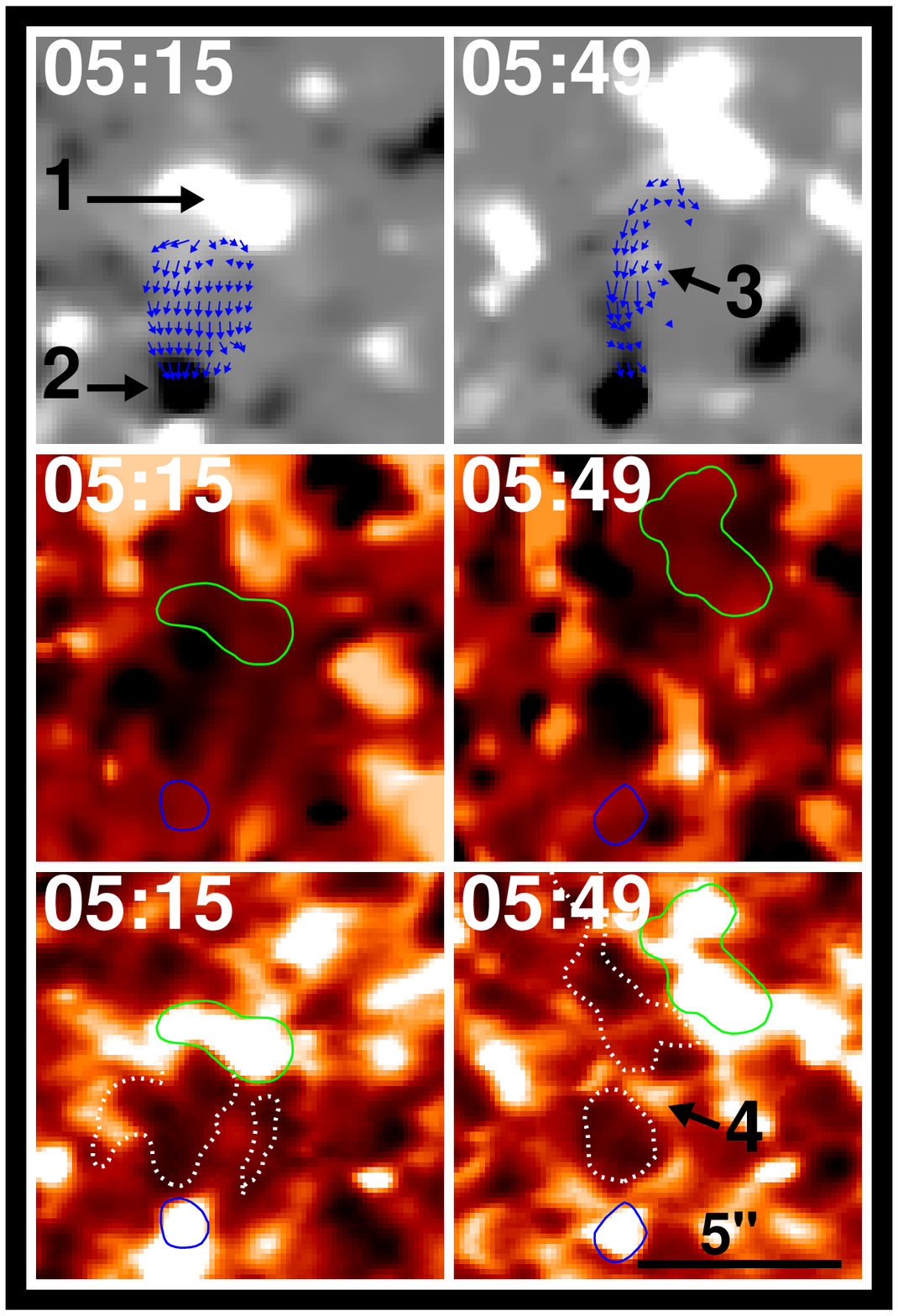}} \quad \subfigure{\includegraphics[bb=145 198
213 632,clip,angle=0,scale=.625]{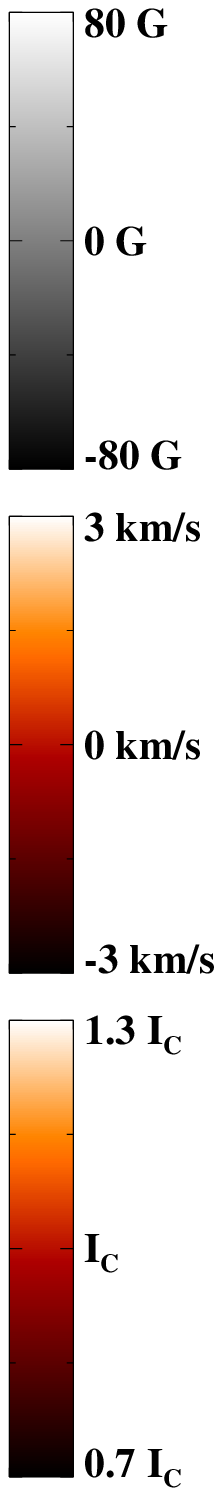}}} \caption{LOS
magnetograms (region ``I" in Fig. 1) in different time (the upper
panels) with an emerging dipole being indicated by arrow ``1"
(positive) and arrow ``2" (negative), corresponding Dopplergrams
(the middle panels) and \ion{Ca}{ii} H images (the bottom panels).
The green and blue curves are contours of the positive ($+$80 G) and
negative ($-$80 G) elements, respectively. The blue arrows show the
transverse fields between the dipolar elements, while the dotted
lines are contours of the Doppler blue shift ($-$2.0 km$/$s). Arrows
``3" and ``4" denote the place of a newly formed magnetic
concentration. The I$_{c}$ (same as in Figs. 4, 6, and 7) marked on
the color bar represents the average value of \ion{Ca}{ii} H
intensities in the FOV of \ion{Ca}{ii} H images. \label{fig3}}
\end{figure}

Magnetic flux emergence is a main form of magnetic field
evolution.
Here, we study an emerging dipole at the CH boundary using the
data from \emph{HINODE}$/$SOT.

Figure 3 shows the region ``I" in Fig. 1 in an expanded view. Two
LOS magnetograms (the upper panels) display an emerging dipole
being indicated by arrows ``1" (positive) and ``2" (negative). The
blue arrows show the transverse magnetic fields between the
dipolar elements, while the dotted lines are contours of the
Doppler blue shift ($-$2.0 km$/$s). We can see that, the
transverse fields between the dipolar elements point from the
positive element to the negative one, and on the narrow zones
where the transverse fields lie, there are strong blue shifts
relative to the surrounding parts. In the bottom panels, the
darker features on the \ion{Ca}{ii} H images mainly coincide with
the stronger blue shift places. Note that the break of the Doppler
structure at 05:49 UT (indicated by arrow ``4") is caused by a
newly emerged flux on the corresponding place (pointed by arrow
``3").

\begin{figure*}
\sidecaption \resizebox{13cm}{!}{\subfigure{\includegraphics
[scale=.5]{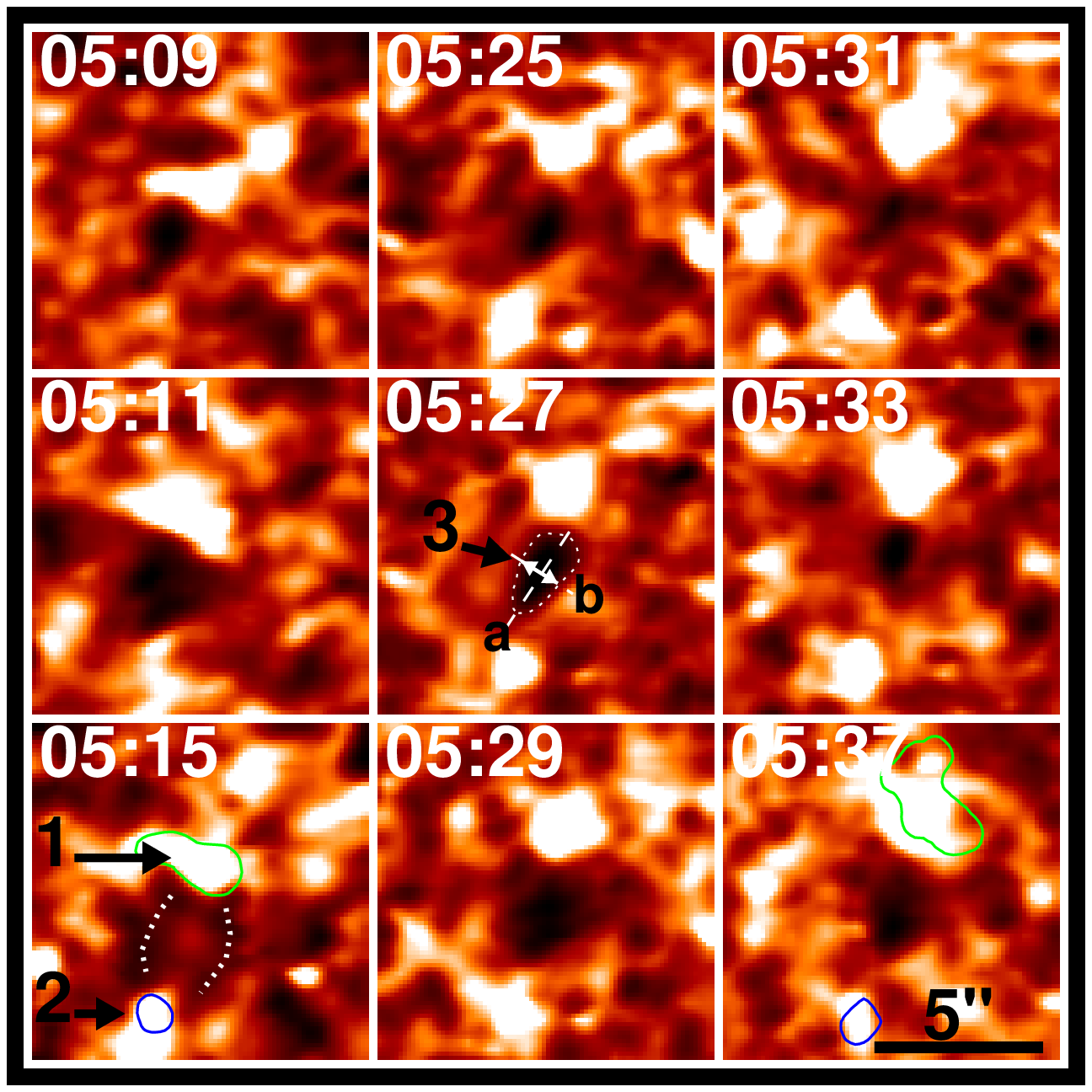}} \quad \subfigure{\includegraphics[bb=145 198
213 632,clip,angle=0,scale=.425]{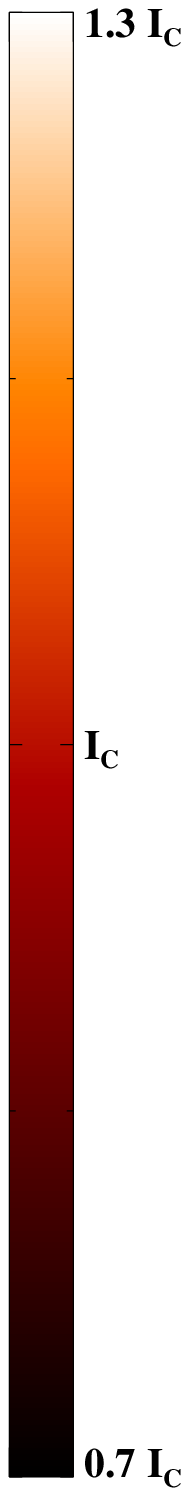}}}
\caption{Sequence of \ion{Ca}{ii} H images showing three developing
processes of three mini AFs led by the emergence of the new dipole
(see Fig. 3). The dipolar elements (indicated by arrows ``1" and
``2") at 05:15 UT and 05:49 UT have been respectively contoured to
the temporally nearest panels. Two thick dotted lines indicate two
segments divided from the initial one. Arrow ``3" denotes the main
body of the AF (described with the thin dotted line). Line ``a" is
drawn along filament, while line ``b" is perpendicular to line ``a"
and put at the midpoint of the intercept of ``a" with the contour.
Bi-head arrow expresses the width of the AF. \label{fig4}}
\end{figure*}

The chromospheric response to the emergence of that dipole is
shown in Fig. 4. The sequence of \ion{Ca}{ii} H images shows the
development process of arch filaments (AFs). The dipolar elements
(indicated by arrows ``1" and ``2") at 05:15 UT and 05:49 UT have
been respectively contoured to the temporally nearest panels.
Three AFs have been observed in turn. Each column shows mainly the
development of one AF.
The first AF appeared at 05:07 UT and then expanded larger. At
05:09 UT, the AF looked like a dark absorption region. This AF
expanded firstly and then divided into two segments. At 05:15 UT,
the two segments connecting two brighter patches were much
obvious. Comparing with SP magnetograms, we noticed that the
bright patches in \ion{Ca}{ii} were co-spatial with the
concentrations of the magnetic field of both polarities. The two
segments persisted for several minutes without obvious change in
size. At 05:23 UT, another similar AF appeared at the same
location as the previous one and began growing. Two minutes later,
the AF was much more obvious seen in the image labeled with
``05:25". Then the AF erupted suddenly (at 05:29 UT) and
dissipated quickly, so that it could not be distinguished from the
background in the next image (at 05:31 UT). Arrow ``3" denotes the
main body of this AF (described with the thin dotted line). In a
similar way to the middle column, the right shows the erupting
process of the last AF. It emerged just following the eruption of
the former one, continued expanding and erupted suddenly at 05:41
UT.

\begin{figure}
\centering \resizebox{8.9cm}{!}{\includegraphics[bb=86 245 486
559,clip,angle=0]{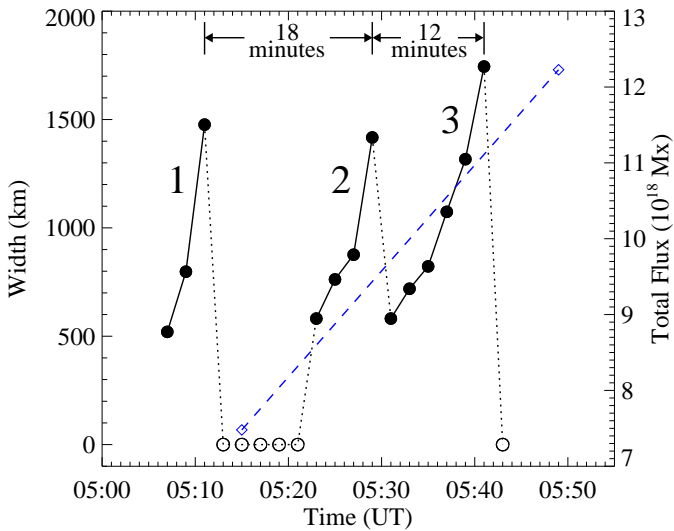}} \caption{Temporal variation of the width
(circle symbols) of the AFs and the total unsigned magnetic flux
(diamonds) of the dipole. The solid line ``1" (``2"$/$``3")
connecting the filled circles represents one visible dividing
(erupting) process, while dotted lines indicate the invisible
stages. When the AF divided or no AF could be distinguished from the
background, the width is set to zero (unfilled circles). The dashed
line represents the change of the total unsigned magnetic
flux.\label{fig5}}
\end{figure}

We measure the width (bi-head arrow in Fig. 4) of the AFs in each
Ca II image and show the temporal evolution of the measured width
with the filled circle symbols in Fig. 5. When the AF divided or
no AF could be distinguished from the background, the width is set
to zero and represented with unfilled circles. The solid line ``1"
(``2"$/$``3") connecting the filled circles represents the visible
width variation of an AF, while dotted lines indicate the
invisible stages. The lifetimes of these three AF are 4, 6, 10
minutes and the two intervals between the three divisions or
eruptions are 18 and 12 minutes, respectively. Also plotted in
this figure is the total unsigned magnetic flux (connected with a
dashed line) of the dipole, increasing from
7.48${\times}$10$^{18}$ Mx to 1.22${\times}$10$^{19}$ Mx in 34
minutes.

\subsection{Magnetic flux emergence in the CH}

\begin{figure}
\centering \resizebox{8.9cm}{!} {\includegraphics[bb=101 233 482
281,clip,angle=0,scale=.425]{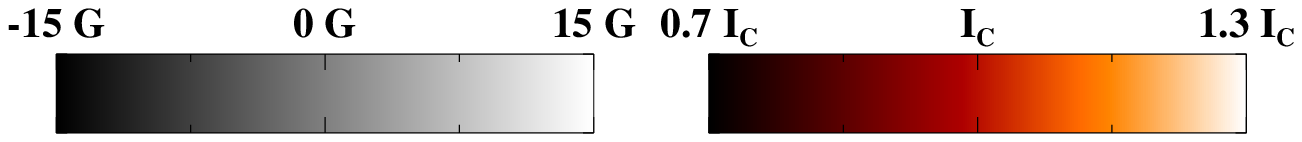}}
\resizebox{8.9cm}{!} {\includegraphics [scale=.5]{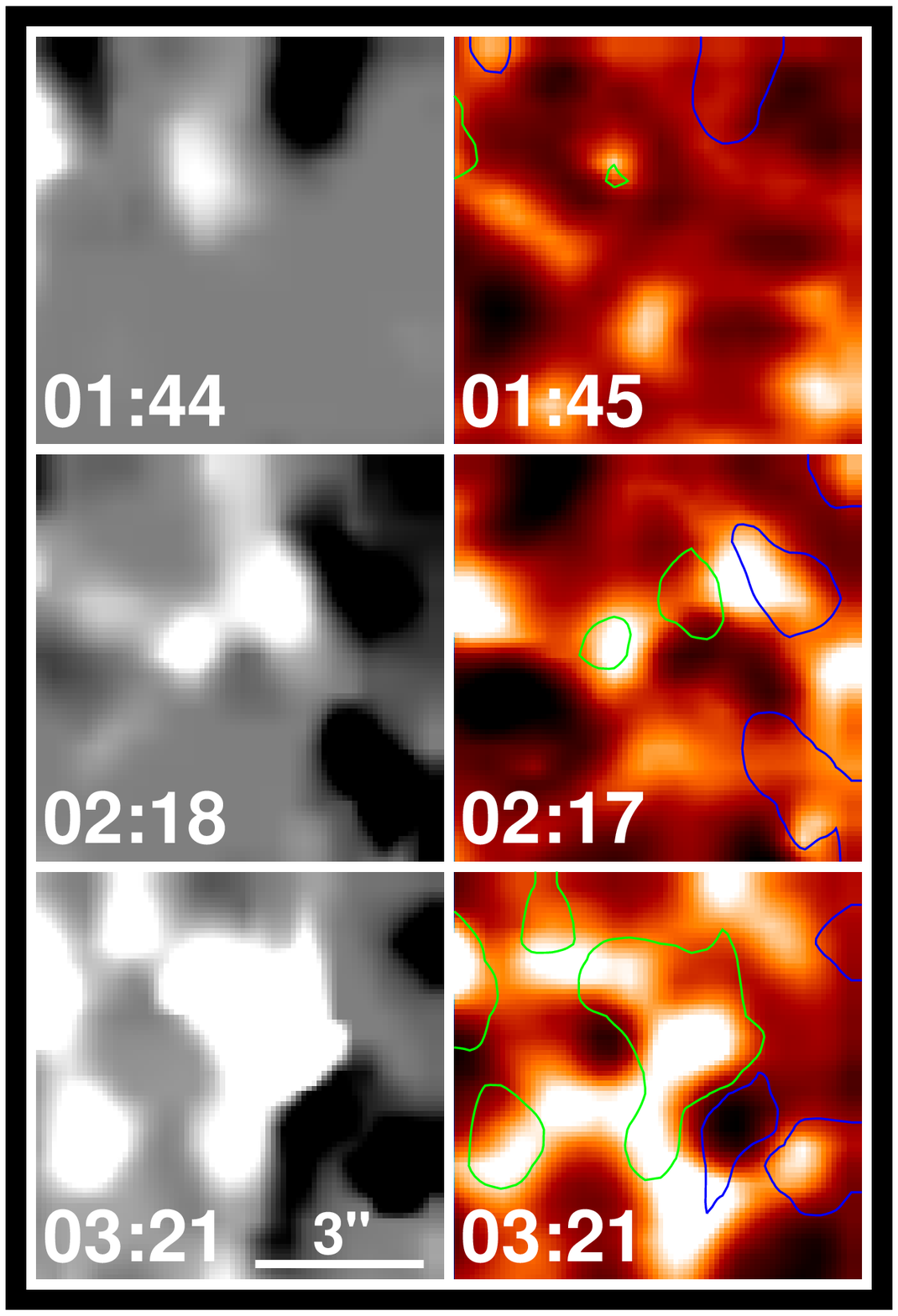}}
\caption{Example of IN flux emergence of mixed-polarity clusters
(region ``II'' in Fig. 1). The left column shows LOS magnetograms,
while the right column consists of \ion{Ca}{ii} H images. Green and
blue contours represent the positive and negative elements at $+$15
G and $-$15 G levels, respectively. \label{fig6}}
\end{figure}

Except for the obvious dipolar emergence, we also find that many
magnetic fluxes emerge in the form of mixed-polarity clusters.
Figure 6 shows an example of IN flux emergence (region ``II'' in
Fig. 1) in the CH. The left column shows LOS magnetograms, while
the right one consists of \ion{Ca}{ii} H images. Green and blue
contours represent respectively the positive and negative magnetic
elements at $+$15 G and $-$15 G levels. The total unsigned
magnetic flux of this region increased from
5.17${\times}$10$^{17}$ Mx (01:44 UT) to 1.24${\times}$10$^{18}$
Mx (02:18 UT), then reached 2.47${\times}$10$^{18}$ Mx (03:21 UT).
On the other hand, we calculate the brightness of the \ion{Ca}{ii}
H line images and find it increased about 15\% at 03:21 UT
compared to that at 01:45 UT, exhibiting an increasing trend.

\subsection{Magnetic flux cancellation at the CH boundary}

\begin{figure}
\centering
\resizebox{8.9cm}{!} {\includegraphics[bb=101 233 498
281,clip,angle=0,scale=.425]{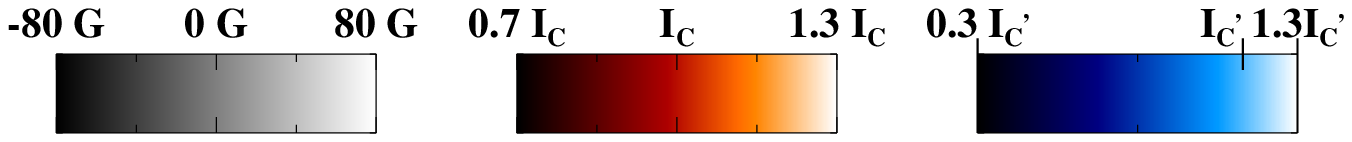}}
\resizebox{8.9cm}{!} {\includegraphics [scale=.5]{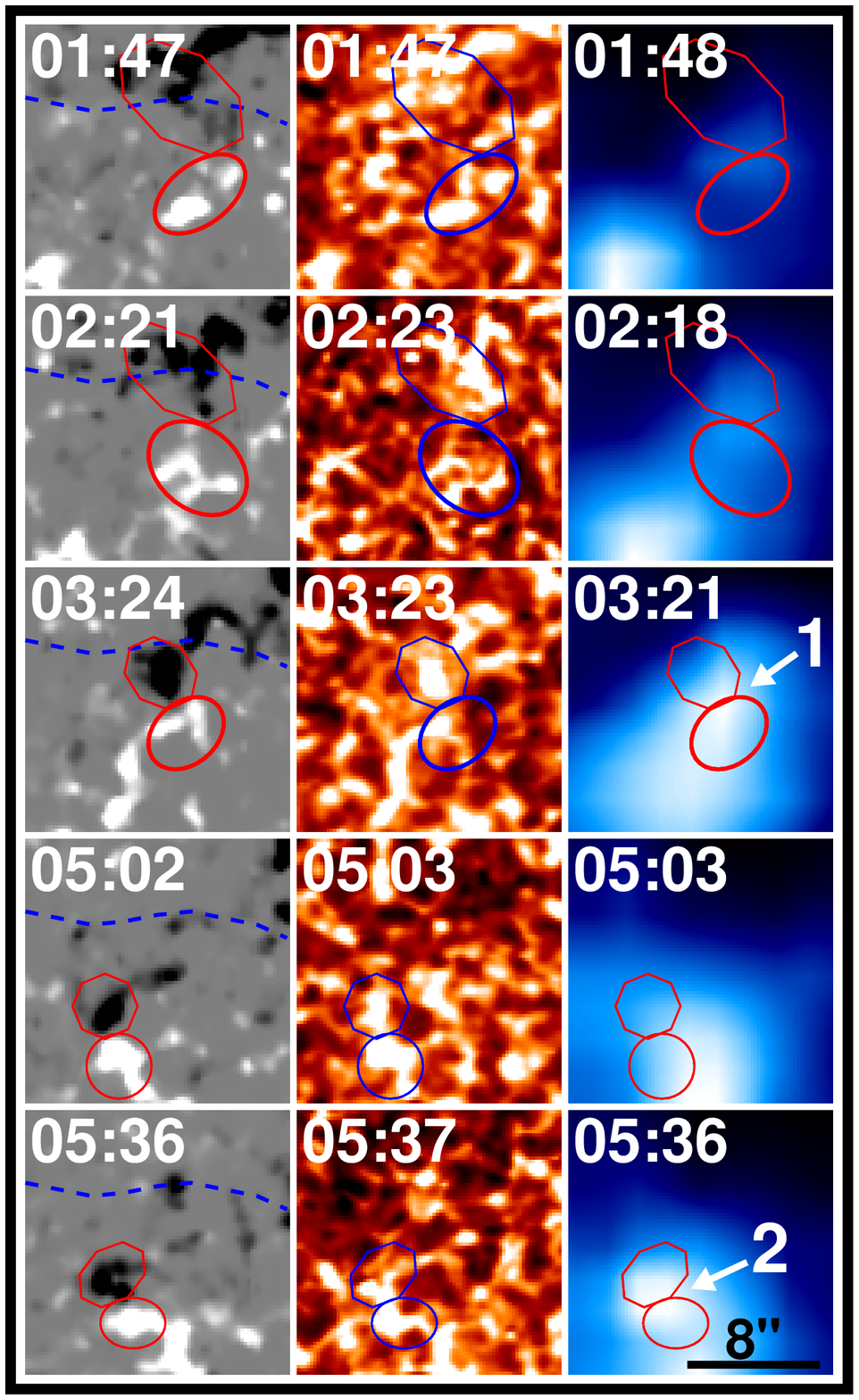}}
\caption{Example of NT flux cancellation locating at the CH boundary
(region ``III'' in Fig. 1). Time sequence of LOS magnetograms (left
column) shows the cancellations between one group of negative
elements (outlined with octagons) and two groups of positive
elements (outlined respectively with ellipses of thick and thin
line) during their evolution. Dashed lines represent the boundary of
the CH. Middle column images are the corresponding response of
\ion{Ca}{ii} H line, whereas right column ones 171 {\AA} line. Note
that the adjacent areas of the octagons and the ellipses are the
places where the cancellations occur. Arrows ``1" and ``2" indicate
the areas where much stronger cancellations take place. The I$_{c'}$
represents the average value of 171 {\AA} intensities in the
full-disk of 171 {\AA} images. \label{fig7}}
\end{figure}

Magnetic flux cancellation is another primary form of magnetic
field evolution. In region ``III" (see Fig. 1) that locates at the
boundary of the CH, we observe the flux cancellations that
occurred between one group of negative NT elements and two groups
of positive ones in turn, as shown in Fig. 7. Left column shows a
time sequence of magnetic field evolution. Dashed lines represent
the CH boundary. The negative elements (outlined with octagons),
initially locating in the CH, moved toward the boundary and
cancelled partially with the positive elements (outlined with
ellipses of thick lines) outside the CH. At 05:02 UT, the negative
elements had become much smaller. They encountered another
positive cluster (outlined with ellipses of thin lines) and
interacted with it. Middle column images show the corresponding
response of \ion{Ca}{ii} H line, whereas right column 171 {\AA}
line. We can see that the bright points of \ion{Ca}{ii} H are
coincided well with the magnetic NT elements in locations, while
the brighter regions seen in the 171 $\AA$ images only appear to
coincide with the underlying magnetic elements when they are
interacting.

As indicated by arrows ``1" and ``2" at 03:21 UT and 05:36 UT,
much stronger cancellations took place with more obvious
brightening of the upper atmosphere. The 171 $\AA$ brightness at
the cancelling areas kept increasing during the two cancelling
courses. At 03:21 UT and 05:36 UT, it increased respectively about
10\% and 8\% than that of pre-cancellation phase.

\subsection{Magnetic flux distribution in the CH}

\begin{figure*}
\sidecaption \resizebox{14cm}{!}{\includegraphics[bb=47 310 555
747,clip,angle=0]{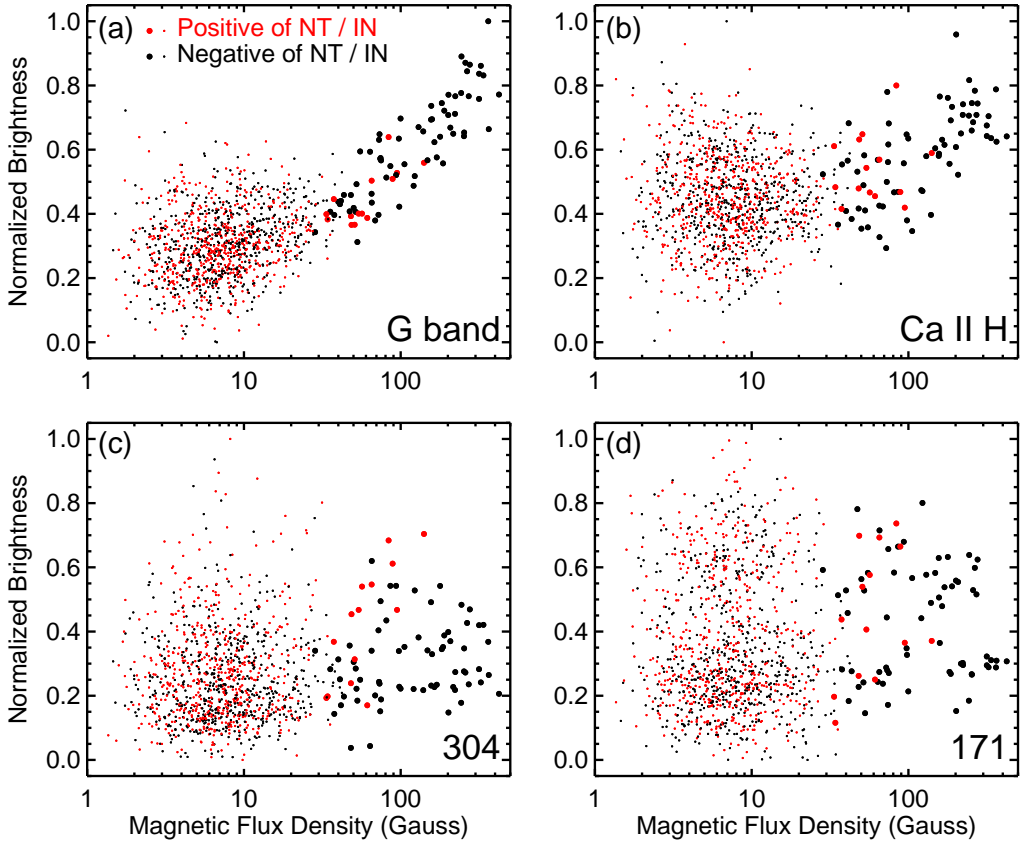}} \caption{Relationship between the
normalized brightness of the images of different wavelengths and the
absolute values of the magnetic flux densities. The big (small) red
and black dots represent the positive and negative data of the NT
(IN) elements, respectively.\label{fig8}}
\end{figure*}

Besides the above case studies of the atmospheric response to the
magnetic field evolution, we have investigated the relations
between the distribution of magnetic flux and brightness at
different atmospheric layers in the CH (only the common part of
the SP and BFI FOVs). We measure the magnetic flux of each NT and
IN element in the LOS magnetogram obtained from 04:51 UT to 05:23
UT and the brightness of the corresponding images of several
spectral lines. In order to compare the results of both negative
and positive elements, two kinds of them are measured separately.
For the NT fields, 66 negative and 14 positive elements are
identified and measured, and the total magnetic flux is
$-$1.85$\times$10$^{20}$ Mx and 0.21$\times$10$^{20}$ Mx,
respectively. We also measure 593 negative and 571 positive IN
elements, and the total flux is $-$0.97$\times$10$^{20}$ Mx and
0.78$\times$10$^{20}$ Mx, respectively. Then we deduce the average
magnetic flux density and brightness of each element.

Figure 8 shows the relationship between the normalized brightness
and the magnetic flux densities. The big (small) red and black
dots represent the positive and negative NT (IN) elements,
respectively. Figure 8a demonstrates that there exists a high
positive correlation between the G band brightness and the
magnetic flux density for the NT elements. If two polarities are
synthetically considered, the linear correlation for the NT will
be as high as 0.85. We also calculate the linear correlation
coefficient for the IN and find it is only 0.27. Figure 8b
exhibits a linear correlation for the NT of the \ion{Ca}{ii} H,
only distinguished from that of G band with a lower correlation
coefficient (0.58). But neither the positive nor the negative IN
has this type of trend. Then Fig. 8c (8d) displays no correlation
between the 304 {\AA} (171 {\AA}) brightness and the magnetic flux
densities for both the NT and the IN. The results of 195 {\AA} and
284 {\AA} are similar to Figs. 8c$-$d.


\section{Conclusions and discussion}

We have investigated an equatorial CH and its boundary region
observed simultaneously by \emph{HINODE} and \emph{STEREO} on July
27, 2007. With the help of the vector magnetic fields and the
Dopplergrams, we study an emerging dipole locating at the CH
boundary. Three AFs accompanying the dipole appeared and expanded
in turn, observed with \ion{Ca}{ii} H line. The first AF divided
into two segments in its late stage. The second and third AFs
erupted and diffused with an extremely rapid speed in their late
stages. The lifetimes of these three AFs are 4, 6, 10 minutes, and
the two intervals between the three divisions or eruptions are 18
and 12 minutes, respectively. We display an example of
mixed-polarity flux emergence of IN fields within the CH and
present the corresponding chromospheric response. With the
increase of the integrated magnetic flux, the brightness of the
\ion{Ca}{ii} H images exhibits an increasing trend. In addition,
we also study magnetic flux cancellations of NT fields locating at
the CH boundary and present the obvious chromospheric and coronal
response. By examining the magnetic NT and IN elements and the
response of different atmospheric layers, we find there exists a
good positive linear correlation (correlation coefficient 0.8)
between the G band brightness and the magnetic flux density for
the NT. The correlation coefficient for the IN of G band is much
lower, only 0.27. We also obtain good linear correlation (
correlation coefficient 0.58) for the NT of \ion{Ca}{ii} H line.

\begin{figure}
\centering
\resizebox{8.9cm}{!}{\includegraphics{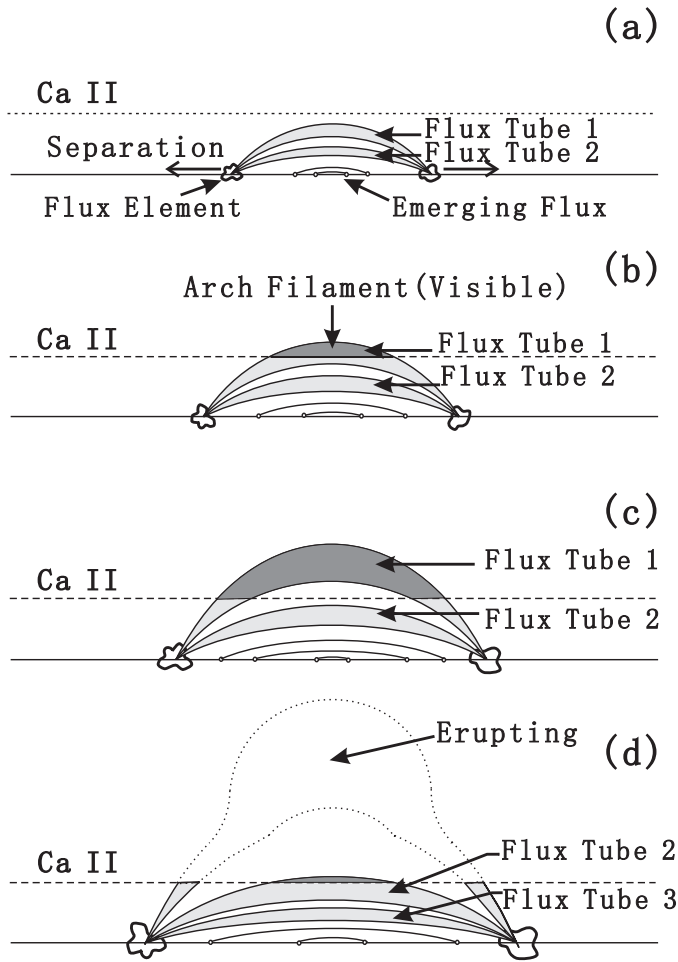}} \caption{Cartoons
illustrating the multi-erupting process of the small-scale AFS.
Dashed lines indicate the lower limit where the \ion{Ca}{ii} H line
is formed. In phase ``a", there is no visible AF at the moment
because all the magnetic flux tubes are too low to be observed with
\ion{Ca}{ii} H line. Then the tubes rise and part of flux tube 1
begins to be observed as AF (darker part), as shown in phase ``b".
From phase ``b" to phase ``c", tube 1 continues to rise and expand,
so the AF apparently becomes bigger. In the late stage (phase ``d"),
the AF erupts suddenly due to some physical mechanism and gets
optically thin and invisible instantly within the visible region of
the \ion{Ca}{ii} H line, following which another developing and
erupting process for flux tube 2 begins.\label{fig9}}
\end{figure}

In order to illustrate the intermittent appearance and eruptions
of the AFs, a series of cartoons (see Fig. 9) are sketched out,
based on the previous model provided by Frazier (1972). The
process of the multi-eruptions can be decomposed into four phases
(Figs. 9a$-$d). Dashed lines indicate the lower limit where the
\ion{Ca}{ii} H line is formed. In Fig. 9a, there is no visible AF
at the moment because all the magnetic flux tubes are too low to
be observed with \ion{Ca}{ii} H line. Then the tubes rise and part
of flux tube ``1" begins to be observed as an AF (darker part)
(Fig. 9b). Tube ``1" continues to rise and expand, so the AF
apparently becomes bigger (Fig. 9c). In the late stage (Fig. 9d),
the AF erupts suddenly due to some physical mechanism and gets
optically thin and invisible instantly within the visible region
of the \ion{Ca}{ii} H line, following which another AF develops.
In each phase, the dipolar footpoints separate persistently,
accompanied with new flux emerging.

Yamauchi et al. (2005) have studied the eruptions of
mini-filaments, running nearly parallel to the neutral lines and
crossing them with small angles, in a CH. A filament, exhibited in
an example, erupted and rose to a high level where its top was
invisible. Then a new visible filament formed along the initial
path of the erupted one. Similarly, rising rapidly to an excessive
height above the layer where the \ion{Ca}{ii} H line is formed may
be the cause of the invisibility of AFs in their late stages
reported in this paper, no matter whether they are optically thin
or not.

CH boundaries separate the CHs from the surrounding quiet regions
and play crucial roles in the evolution of the CHs. Kahler \&
Hudson (2002; see also Dahlburg \& Einaudi 2003) have researched
the magnetic morphology of CH boundaries. These studies are
helpful to understand the rigid rotation which is most likely
correlated with the processes occurring at the boundaries (Nash et
al. 1988; Wang \& Sheeley 1990, 2004; Fisk et al. 1999; Fisk \&
Schwadron 2001). In order to maintain the integrity of the CHs,
magnetic reconnection must occur continuously at the boundaries
(Wang et al. 1996; Kahler \& Hudson 2002). Wang et al. (1996) and
Schwadron et al. (1999) suggested that there exists magnetic
reconnection between the open field lines and the adjacent closed
ones along the boundaries. The evidence for such reconnection was
revealed with the existence of bidirectional jets by Madjarska et
al. (2004) for the first time. Then more direct evidence was given
by Raju et al. (2005) and Baker et al. (2007).
When magnetic reconnection occurs, the magnetic field is
restructured accompanied with energy release (e.g. bright point
appears in EUV image); meanwhile small loop forms and submerges
leading to an observational phenomenon $-$ magnetic flux
cancellation, just as shown in Fig. 7 (see also Wang \& Shi 1993;
Zhang et al. 2001, 2007).

Our result that there are good positive linear correlations
between the NT magnetic flux densities and the brightness of both
G band and \ion{Ca}{ii} H in the CH is consistent with the
research of Zirin (1988) and Nindos \& Zirin (1998) in the QS. The
lower correlation for the NT of \ion{Ca}{ii} H than of G band can
be considered to be caused by the fact that, when the flux tubes
extend to the chromosphere, they expand, becoming larger than in
the photosphere. With the increase of the height, the flux tubes
enlarge quite a lot and most of them change their directions.
Especially in the corona, the average expansion factor (ratio of
the maximum to the minimum of the flux tube area) is 28 ${\pm}$ 11
(Wiegelmann et al. 2005). The decline and expansion of the flux
tubes can be used to interpret why there is no correlation in the
transition region and the corona (exhibited by Figs. 8c and 8d),
although the relatively lower tempo-spatial resolution of the
images from \emph{STEREO} may be one cause.

The large part of scatter of the data points for the IN of
\ion{Ca}{ii} H in Fig. 8b is mainly due to the 3-min oscillations,
which does not permit us to obtain a correlation between the
brightness and magnetic fields even if it exists on the Sun
(Sivaraman et al. 2000). The low correlation coefficient (0.27)
for the IN of G band in Fig. 8a also may be affected by the 3-min
oscillations. Another reason for the lack of obvious linear
correlation for both G band and \ion{Ca}{ii} H IN may be that the
cadence of G band and Ca images is not high enough. The average
lifetime of the IN elements is only about 4.84 minutes (Zhou, G.
private communication), so lots of the magnetic IN elements have
evolved quite a lot within one minute. The same reasons can also
explain why there is only partial coincidence between the magnetic
elements and the bright features shown in Fig. 6. While the poor
correlations for the IN of 304 {\AA}, 171 {\AA}, 195 {\AA} and 284
{\AA} lines are mainly due to the lower tempo-spatial resolution
of the images and the incapability of the IN flux tubes to extend
into the transition region and the corona (Wiegelmann \& Solanki
2004; Wiegelmann et al. 2005).

\begin{acknowledgements}
We are grateful to the \emph{HINODE}, \emph{STEREO} and
\emph{SOHO} teams for providing the data. \emph{HINODE} is a
Japanese mission developed and launched by ISAS$/$JAXA, with NAOJ
as domestic partner and NASA and STFC (UK) as international
partners. It is operated by these agencies in co-operation with
ESA and NSC (Norway). This work is supported by the National
Natural Science Foundations of China (G40674081, 40890161,
10573025, 10703007, and 10733020), the CAS Project KJCX2-YW-T04,
and the National Basic Research Program of China under grant
G2006CB806303.
\end{acknowledgements}

\clearpage

\end{document}